\documentclass[fleqn,twoside,narrowdisplay,final]{article}
\usepackage[fleqn]{amsmath}
\usepackage{amsfonts}
\usepackage{amsthm}
\usepackage{espcrc2}
\usepackage{graphicx}
\usepackage[figuresright]{rotating}

\newtheorem{lemma}{Lemma}
\newtheorem*{proposition}{Proposition}

\theoremstyle{remark}
\newtheorem{remark}{Remark}
\newtheorem*{remark*}{Remark}
\newtheorem{example}{Example}

\begin{document}
\title{On the Poisson Approximation to Photon Distribution for Faint Lasers}

\author{Yucheng Hu$^{\rm{a}}$, Xiang Peng$^{\rm{b}}$,
Tiejun Li\address{School of Mathematical Sciences, Peking University, Beijing 100871,
P.R. China}\thanks{Author to whom correspondence should be
addressed. E-mail: hongguo@pku.edu.cn, phone: +86-10-6275-7035, Fax:
+86-10-6275-3208.} and
Hong Guo\address{Key Laboratory for Quantum Information
and Measurements of Ministry of Education, School of Electronics
Engineering and Computer Science, Peking University, Beijing 100871,
P.R. China}\thanks{Author to whom correspondence should be
addressed. E-mail: tieli@pku.edu.cn.}}

\begin{abstract}
It is proved, that for a certain kind of input distribution, the strongly
binomially attenuated photon number distribution can well be approximated by a
Poisson distribution. This explains why we can adopt poissonian distribution as the
photon number statistics for faint lasers. The error of such an approximation is
quantitatively estimated. Numerical tests are carried out, which coincide with our
theoretical estimations. This work lays a sound mathematical
foundation for the well-known intuitive idea which has been widely used in quantum cryptography.
\end{abstract}


\maketitle

\section{Introduction}
The security of Quantum Key Distribution (QKD) is based on the
non-cloning principle of an unknown quantum state \cite{bnc}. In
the implement of QKD based on BB84 protocol \cite{bb84}, one expect that each
pulse contains only one photon. If not,
the eavesdropper can acquire information
using beamsplitter attack \cite{lh} without exposing its
existence. However, since an ideal single photon state is difficult to prepare,
practically, faint laser pulse with ultra-low mean
photon number is used as a convenient realization of
pseudo-single photon source \cite{Gisin}.

By letting a laser source pass through a strong attenuator we get
faint laser pulse. For security concern the mean photon number in each
faint laser pulse is kept very small (about 0.1).
But still, there is a small probability of having more than one
photon in each pulse. A precise estimation of this unwelcome
probability is crucial for security analysis in QKD systems
\cite{lh}.

In the literature, the photon number in faint laser is
treated as Poissonian distributed. It is all right if the input laser
before attenuation is Poisson. However, practically we may have input laser
whose photon number statistics is not Poisson \cite{np}. If this laser is
used as input, the attenuated laser may not be Poisson either. But there is a
common belief that no matter what distribution the input laser is,
if we attenuate it into a faint laser with sufficient small
mean photon number, then Poisson distribution would be a good approximation
of photon number distribution in the faint laser pulse. So far, however, this claim has not been
mathematical rigorously proved, which is the motivation of this work.

We proved, that for a certain kind of probability distribution, after
the binomial decay transformation, which is a mathematical
description of laser attenuation \cite{Marcuss}, the decayed
distribution can well be approximated by a Poisson distribution
provided that the expectation of the decayed distribution is
sufficient small. It gives a theoretical validation of the above
claim, i.e.,  generally we can use a Poisson distribution to approximate
the photon distribution in faint laser and the error of the
approximation could be neglected.

\section{Preliminary}
Consider $N$ independent particles (photons in laser pulses) passing
through an attenuator. Each particle has a probability of $\eta$
($0\le\eta\le 1$) to penetrate the attenuator. We define
$X$ to be the number of particles before decay (Input); and
$X_\eta$ to be the number of particles after decay (Output).
$X$ and $X_\eta$ are random variables taking values in the
natural number system $\mathbb{N}$, and their probability mass
functions (PMF) are $P(N)$ and $P_\eta(n)$.

We can connect $P(n)$ and $P_\eta(n)$ via the \emph{binomial decay
transformation} \cite{Marcuss}
 \begin{equation}\label{2.1}
 P_{\eta}(n) = \sum_{N=n}^{\infty}
 \binom{N}{n}
 \eta^{n}(1-\eta)^{N-n}P(N).
 \end{equation}
 It is easy to check that
\begin{equation*}
P_{\eta}(n)\ge 0, \ \forall n\in \mathbb{N},
\end{equation*}
 and
\begin{equation*}
\sum_{n=0}^{\infty}P_{\eta}(n) = \sum_{N=0}^{\infty}P(N) = 1.
\end{equation*}
 So $P_{\eta}$ is indeed a PMF.

The binomial decay transformation establishes a relation
between photon number distribution before and after the attenuation.
The remaining of this section gives some properties of it.

\begin{lemma}\label{lm1}
 Let $E(X)$ and $E(X_{\eta})$ denote the
expectation of the discrete random variables $X$ and $X_{\eta}$,
respectively. Then $E(X_{\eta}) = \eta E(X)$.
\end{lemma}
\begin{proof}
By the definition of expectation and Eq.~(\ref{2.1}),
\begin{eqnarray*}
E(X_{\eta}) &=& \sum_{n=0}^{\infty} n P_\eta(n) \\
            &=& \sum_{n=0}^{\infty} n \sum_{N=n}^{\infty}
            \binom{N}{n}
          \eta^{n}(1-\eta)^{N-n}P(N)\\
          &=& \sum_{N=0}^{\infty}P(N)\sum_{n=0}^{N} n
          \binom{N}{n}\eta^{n}(1-\eta)^{N-n}\\
            &=& \sum_{N=0}^{\infty}\eta N P(N) = \eta E(X).
\end{eqnarray*}
\end{proof}
\noindent Define $P^{\lambda}(n)$ the Poisson PMF with parameter $\lambda$,
i.e., $P^{\lambda}(n) = e^{-\lambda} \lambda^n/n!, n\in \mathbb{N}$ \cite{Gardiner,durrett}.
We will show that the binomial decay transformation preserves the
Poisson character.
\begin{lemma}\label{lm2}
Suppose $P(N)=P^{\mu}(N)$, then we have $P_\eta(n)=P^{\eta \mu}(n).$
\end{lemma}
\begin{proof}
From Eq.~(\ref{2.1}),
\begin{eqnarray*}
P_\eta(n) &=& \sum_{N=n}^{\infty}
\binom{N}{n}\eta^{n}(1-\eta)^{N-n}P^\mu(N)\\
          &=& \frac{(\eta\mu)^n}{n!} \sum_{N=n}^{\infty}
\frac{e^{-\mu}}{(N-n)!}\left[\mu(1-\eta)\right]^{N-n} \\
            &=& \frac{(\eta\mu)^n}{n!}e^{-\eta\mu}.
\end{eqnarray*}
\end{proof}

\section{Poisson Approximation}
From Lemma \ref{lm2} and Lemma \ref{lm1} we know that if $P(N)$,
the photon number distribution in the laser
before attenuation, is Poisson with parameter $\mu$,
then the decayed distribution, $P_\eta(n)$, is also a Poisson whose parameter is $\eta \mu$,
with $\eta$ being the attenuating coefficient.
However, practically $P(N)$ may not be a Poisson
distribution \cite{np}. If so, $P_\eta(n)$ would not be a Poisson
distribution. Nevertheless, in QKD the faint
laser is treated as Poisson distributed.
The reason of doing this is based on
the common belief that any input distribution
would reduce to Poisson distribution provided that
the attenuation is strong enough. Next we
justify it quantitatively.

\begin{proposition}\label{mp}
In the case of faint laser, the decayed distribution can be
approximated as a Poisson distribution,
\begin{equation*}
P_\eta(n)\approx P^{\lambda}(n),\ \ n\in \mathbb{N},
\end{equation*}
where
\begin{equation*}
\lambda  = E(X_{\eta}) = \eta E(X)\ll 1.
\end{equation*}
More concretely, if we expand $P^{\lambda}(n)$ into Taylor series of
$\lambda$:
\begin{subequations}
\begin{eqnarray}
P^{\lambda}(0) &=& 1- \lambda + \frac{\lambda^2}{2} - O(\lambda^3),\label{pl0}\\
P^{\lambda}(1) &=& \lambda - \lambda^2 + O(\lambda^3),\\
P^{\lambda}(2) &=& \frac{\lambda^2}{2} - O(\lambda^3),\\
P^{\lambda}(n) &=& O(\lambda^3), n \ge 3.\label{pln}
\end{eqnarray}
\end{subequations}
Then we have
\begin{subequations}
\begin{equation}
P_{\eta}(0) = 1- \lambda + \frac{\lambda^2}{2} + C(X)\lambda^2 - D_0(X)\lambda^3,\label{pe0}
\end{equation}
\begin{equation}
P_{\eta}(1) = \lambda - \lambda^2 - 2C(X)\lambda^2 +  D_1(X)\lambda^3,\label{pe1}
\end{equation}
\begin{equation}
P_{\eta}(2) = \frac{\lambda^2}{2} + C(X)\lambda^2 - D_2(X)\lambda^3, \label{pe2}
\end{equation}
\begin{equation}
\sum_{n=3}^{\infty}P_{\eta}(n) =  \left[ D_0(X) + D_2(X) - D_1(X) \right] \lambda^3, \label{pen}
\end{equation}
\end{subequations}
where
\begin{equation*}
C(X) = \frac{Var(X) - E(X)}{2E(X)^2}
\end{equation*}
and
\begin{equation*}
0 \le D_i(X) \le D(X) = \frac{M(X)}{E(X)^3}, \ (i=0,1,2).
\end{equation*}
$M(X)=E\left[X(X-1)(X-2)\right]$ is the 3rd factorial moment of $X$.
\end{proposition}

\begin{proof}
The generating function of $X$ is
\begin{equation*}
G(z) = \sum_{N=0}^{\infty}P(N)z^N, z \in \mathbb{R}.
\end{equation*}
Taking the $n$-th order derivatives of $G(z)$ with respect to $z$ yields,
\begin{equation*}
G^{(n)}(z) = \sum_{N=n}^{\infty}N(N-1)\cdots(N-n+1)P(N)z^{N-n}.
\end{equation*}
Let $z=1$, one has,
\begin{equation*}
G^{(n)}(1) = E\left[\frac{X!}{(X-n)!}\right].
\end{equation*}
For $n=0,1,2,3$ we have
\begin{eqnarray*}
G(1) &=& 1,\\
G'(1) &=& E(X),\\
G''(1) &=& Var(X) + [E(X)]^2 - E(X),\\
G'''(1) &=& E\left[X(X-1)(X-2)\right] = M(X).
\end{eqnarray*}
From Eq.~(\ref{2.1}),
\begin{equation*}
P_\eta (n)=\sum_{N=n}^{\infty} \binom{N}{n}\eta^{n}(1-\eta)^{N-n}P(N)
\end{equation*}
\begin{equation*}
=\frac{\eta^n}{n!}\sum_{N=n}^{\infty} N(N-1)\cdots(N-n+1)(1-\eta)^{N-n}P(N)
\end{equation*}
\begin{equation*}
=\frac{\eta^n}{n!}G^{(n)}(1-\eta).
\end{equation*}
Expanding $G^{(n)}(1-\eta)$ into Taylor serials, in the case of $n=0$, one has,
\begin{equation*}
P_\eta (0) = G(1) - \eta G'(1) + \frac{\eta^2}{2}G''(1) - \frac{\eta^3}{6}G'''(1-\theta_0 \eta)
\end{equation*}
\begin{equation*}
= 1 - \lambda + \frac{\lambda^2}{2} +
\frac{Var(X) - E(X)}{2E(X)^2}\lambda^2 -
\frac{G'''(1-\theta_0 \eta)}{6E(X)^3}\lambda^3
\end{equation*}
\begin{equation*}
= 1 - \lambda + \frac{\lambda^2}{2} + C(X)\lambda^2 -
D_0(X)\lambda^3,
\end{equation*}
where $\theta_0 \in [0,1]$ and
\begin{equation*} 0 \le
D_0(X)=\frac{G'''(1-\theta_0 \eta)}{6E(X)^3} \le \frac{M(X)}{E(X)^3} = D(X).
\end{equation*}
Analogously, Eq.~(\ref{pe1}) and Eq.~(\ref{pe2}) can be derived.
Finally, apply the relation
\begin{equation*}
\sum_{n=3}^{\infty}P_{\eta}(n) = 1 - \left[P_\eta(0)+P_\eta(1)+P_\eta(2)\right],
\end{equation*}
one yields Eq.~(\ref{pen}).
\end{proof}

\begin{remark}
If $C(X)$ and $D(X)$ is not too big,
$\lambda^2 C(X)$ and $\lambda^3 D(X)$ can be ignored when
$\lambda\ll 1$. By comparing Eqs.~(\ref{pl0})-(\ref{pln}) with
Eqs.~(\ref{pe0})-(\ref{pen}) we can see that $P_\eta(n)$ is well
approximated by the Poisson distribution $P^\lambda(n)$.
\end{remark}

\begin{remark}
The approximation of Poisson distribution is an asymptotic result in
the limit $\lambda\rightarrow 0$. If $\lambda$ is large, this
approximation will be broken, as can be found from the numerical
Example \ref{exam1} in the following.
\end{remark}

\begin{remark}
The approximation error $\Delta(n) = P_\eta(n) - P^\lambda(n)$ can
be written as
\begin{subequations}
\begin{eqnarray}
\Delta(0) &=& \lambda^2C(X) + O(\lambda^3),\label{e1} \\
\Delta(1) &=& -2\lambda^2C(X) + O(\lambda^3),\\ \label{e2}
\Delta(2) &=& \lambda^2C(X) + O(\lambda^3),\\ \label{e3}
\Delta(n) &=& O(\lambda^3), \ \ n \ge 3. \label{e4}
\end{eqnarray}
\end{subequations}
Here $C(X) = [Var(X) - E(X)]/[2E(X)^2]$
is determined by $P(N)$,
the input distribution only. If $Var(X) = E(X)$, then $C(X) = 0$
and the error decreases to $O(\lambda^3)$. On the other hand, for
some singular input distribution, $C(X)$ is so big that
$P_\eta(n)$ can no longer be approximated by Poisson. In Example
\ref{exam3} we give a typical example that Poisson approximation
fails.
\end{remark}

In QKD we use faint laser to simulate the single photon
source. For security analysis, it is important to estimate
$P_\eta(n>1|n>0)$, the probability that a pulse contains more than
one photon \cite{Gisin}. According to our estimation,
\begin{eqnarray*}
P_\eta(n>1|n>0) &=& \frac{1-P_\eta(0)-P_\eta(1)}{1-P_\eta(0)} \\
& \approx & \frac{\frac{\lambda^2}{2} +
C(X)\lambda^2}{\lambda-\frac{\lambda^2}{2} - C(X)\lambda^2}.
\end{eqnarray*}
Here we have neglected the $\lambda^3$ and higher order terms. We
further simplify it by removing the $\lambda^2$ terms in the
denominator, which gives
\begin{equation}\label{risk}
P_\eta(n>1|n>0) \approx \left[\frac{1}{2} + C(X)\right]\lambda.
\end{equation}

If the input distribution is Poisson, then $C(X) =0$.
After we attenuate it to faint laser that contains an
average of 0.1 photon in each pulse,
which means $\lambda=0.1$, we would have $P_\eta(n>1|n>0) = 0.05$, i.e., each
pulse has about $5\%$ chance to contain more than one photon.

From Eq.~(\ref{risk}) we can see that, the risk of a QKD system rises as $C(X)$ grows.
In Example \ref{exam3} we use an ill-shaped input
distribution whose $C(X)=9.11$.
After we attenuate it to $\lambda=0.1$, then $P_\eta(n>1|n>0)
\approx 1$. So for this big a $C(X)$, the QKD system
would be totally unsecured since almost every pulse contains at
least two photons.

On the other hand, if the input distribution satisfies $Var(X) < E(X)$ then
$C(X)<0$. One can expects the attenuated faint laser be more
secure than a Poisson laser because we get a smaller $P_\eta(n>1|n>0)$. The
last row of table \ref{tab1} gives an example of negative $C(X)$.
However, the possible smallest $C(X)$ is
$-\frac{1}{2}{E(X)}^{-1}$. For practical input laser, $E(X)$ is so
big that $-\frac{\lambda}{2}{E(X)}^{-1}$ can be ignored. This
means, for attenuated faint laser, a Poisson distribution is almost as good as
one can expects.
\begin{figure}[h]
\centering
\includegraphics[width=18pc]{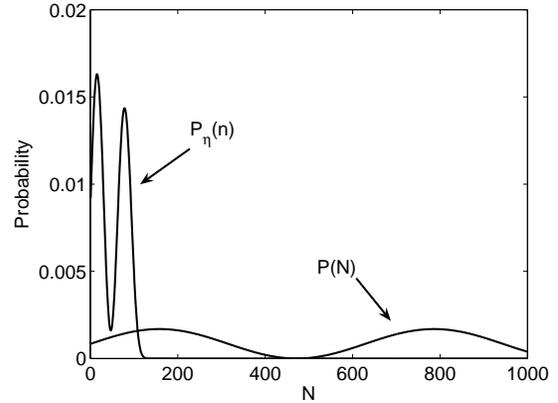}
\caption{If attenuation is not strong enough, the decayed
distribution may not be approximated by Poisson distribution. In
this example the input random variable takes integer values from
$0$ to $1000$, $\eta$ is set to 0.1, and the expectation after
decay is $\lambda = 48.85$.}
        \label{fig1}
\end{figure}

\section{Numerical Examples}
Following we give some numerical simulations and the results coincide with
our theoretical estimations quite well.

\begin{example}\label{exam1} First we show that if the
mean value of the decayed distribution $\lambda = \eta E(X)$ is not
sufficiently small, then in general, approximation using Poisson distribution
fails. We choose a random variable $X$ whose PMF
takes the shape in Figure \ref{fig1} and set $\eta = 0.1$. In this
case $\lambda\approx 48.85$, and it can be observed that $P_\eta(n)$
is far away from Poisson distribution.
\end{example}

\begin{figure}[h]
\centering
\includegraphics[width=18pc]{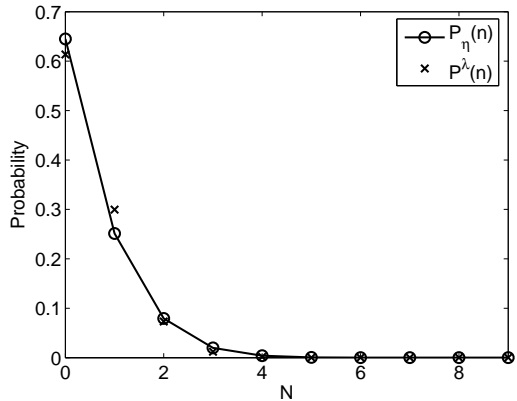}
\caption{Using the same input distribution as Example \ref{exam1},
but we replace the attenuator with a 100 times stronger one with
$\eta = 0.001$ and $\lambda=0.4885$. Only the probability
value of the first few $n$'s are plotted since the others are almost zero.}
        \label{fig2}
\end{figure}

\begin{figure}[h]
\centering
\includegraphics[width=18pc]{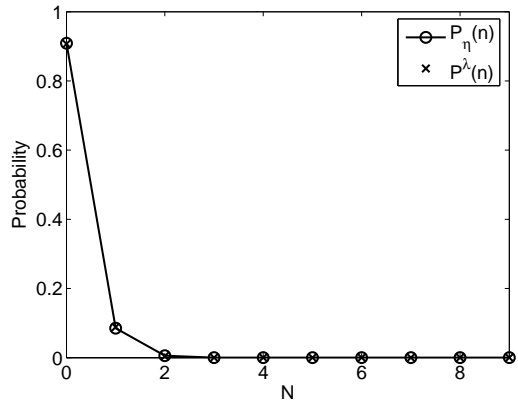}
\caption{Like Example \ref{exam2}, we use a even stronger
attenuator with $\eta = 0.0002$ and $\lambda=0.0977$. }
        \label{fig3}
\end{figure}

\begin{example}\label{exam2}
Using the same input random variable $X$ as Example \ref{exam1}, but
we take $\eta=0.001$, which means that our new attenuator is $100$ times
stronger than the old one. Now we have $\lambda=0.4885$ and the
binomial decayed PMF $P_\eta(n)$ is close to a Poisson distribution,
as indicated in Figure \ref{fig2}. We further decrease
$\eta$ to $0.0002$, then $\lambda=0.0977$ and Figure \ref{fig3}
shows a perfect match between the decayed distribution and Poisson
distribution, which claims strong support for the common
belief that photon number in faint laser pulse can
be treated as Poisson distributed.
\end{example}

\begin{figure}[h]
\centering
\includegraphics[width=18pc]{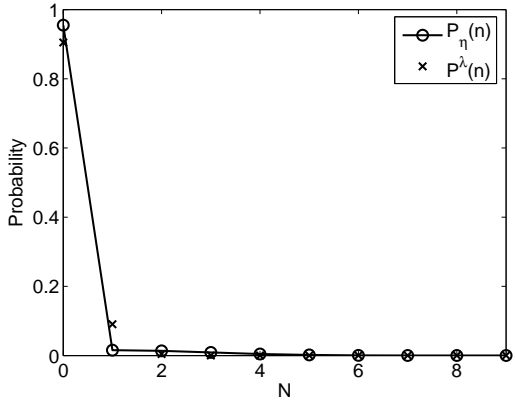}
\caption{A typical example when error is large even if $\lambda$
is small. The input distribution is constructed as $P(X=1) = 0.95$
and $P(X=1001)=0.05$. We decay it to $\lambda=0.1$ and it turns
out that $P^\lambda(n)$ fails to converge to $P_\eta(n)$.}
        \label{fig4}
\end{figure}

\begin{example}\label{exam3}
For some ill-shaped input distributions, $C(X)$ could be
very large. If this happens, even if
$\lambda$ is small, Poisson approximation could still fail. As an
example, we construct a singular input distribution of which $P(X=1)
= 0.95$ and $P(X=1001)=0.05$. In this case, $C(X)=9.11$. If we
attenuate it to $\lambda=0.1$, as Figure \ref{fig4} shows,
$P^\lambda(n)$ fails to converge to $P_\eta(n)$.
\end{example}

To quantitatively test the approximation error, Eqs.~(\ref{e1})-(\ref{e4}),
we choose different input distributions and computer
their attenuated distribution. We adjust the attenuating
coefficient $\eta$ to keep $\lambda = 0.1$. The numerical result,
which is listed in Table \ref{tab1}, supports our theoretical
estimation very well.

\begin{table*}
\caption{\label{tab1}Numerical validation of the approximate error,
Eqs.~(\ref{e1})-(\ref{e4}). It can be observed that
$\Delta(0)\approx \Delta(2)\approx \lambda^2C(X)$, which goes on
well with Eq.~(\ref{e1}) and Eq.~(\ref{e3}); And
$\Delta(1)\approx-2\lambda^2C(X)$, which also agrees with
Eq.~(\ref{e2}).}

\newcommand{\m}{\hphantom{$-$}}
\newcommand{\cc}[1]{\multicolumn{1}{c}{#1}}

\centering
\begin{tabular}{@{}*{6}{l}}
  \hline
 \cc{$\lambda^2C(X)$}& \cc{$\Delta(0)$} & \cc{$\Delta(1)$} & \cc{$\Delta(2)$} & \cc{$\Delta(3)$} & \cc{$\Delta(4)$}  \\
  \hline
 \m0.0045   &\m0.0039   & $-0.0073$  &  \m0.0028 &  \m0.0005  &  \m0.0000  \cr
 \m0.0030   & \m0.0027  & $-0.0050$  & \m0.0020  & \m0.0003   & \m0.0000  \cr
 \m0.0018   & \m0.0017  & $-0.0032$  &  \m0.0013 &   \m0.0002 & \m0.0000  \cr
 \m0.0011   &   \m0.0010& $-0.0019$  &  \m0.0008 &   \m0.0001 &    \m0.0000  \cr
 \m0.0005   &  \m0.0005 &  $-0.0010$ &  \m0.0004 &   \m0.0000 &  \m0.0000  \cr
 $-0.00016$  &  $-0.00014$&  \m0.00027& $-0.00011$ &  $-0.00001$ & $-0.00000$  \cr
  \hline
\end{tabular}
\end{table*}

\section*{Acknowledgement}

This work is partially supported by National Science Foundation of China
(Grant No. 10474004, No. 10401004
and No. 20490222), National Basic Research Program
(Grant 2005CB321704) and DAAD exchange program:
D/05/06972 Projektbezogener
Personenaustausch mit China
(Germany/China Joint Research Program).


\end{document}